\newcounter{myctr}
\def\myitem{\refstepcounter{myctr}\bibfont\noindent\ifnum\themyctr>9\else\phantom{0}\fi\hangindent17pt\themyctr.\enskip}

\documentclass{ws-ijqi}

\newcommand{\be}{\begin{equation}}
\newcommand{\ee}{\end{equation}}
\newcommand{\ba}{\begin{eqnarray}}
\newcommand{\ea}{\end{eqnarray}}
\newcommand{\nn}{\nonumber \\}

\begin{document}

\markboth{A.T.S.~WAN, M.H.S.~AMIN, and S.X.~WANG} {LANDAU-ZENER
TRANSITIONS IN THE PRESENCE OF SPIN ENVIRONMENT}

\catchline{}{}{}{}{}

\title{LANDAU-ZENER TRANSITIONS \\ IN THE PRESENCE OF SPIN ENVIRONMENT}

\author{ANDY T.~S.~WAN\footnote{Currently at UBC mathematics department.} ,
 M.~H.~S.~AMIN, and SHANNON X.~WANG\footnote{Currently at MIT physics
department.}}

\address{D-Wave Systems Inc., 100-4401 Still Creek Drive, \\
Burnaby, B.C., V5C 6G9, Canada}

\maketitle


\begin{abstract}
We study the effect of an environment consisting of noninteracting
two level systems on Landau-Zener transitions with an interest on
the performance of an adiabatic quantum computer. We show that if
the environment is initially at zero temperature, it does not affect
the transition probability. An excited environment, however, will
always increase the probability of making a transition out of the
ground state. For the case of equal intermediate gaps, we find an
analytical upper bound for the transition probability in the limit
of large number of environmental spins. We show that such an
environment will only suppress the probability of success for
adiabatic quantum computation by at most a factor close to 1/2.
\end{abstract}


\section{Introduction}

Theories of open quantum systems have gained renewed attention in
recent years, thanks to their important role in quantum computation.
Such theories usually study quantum evolution of a system in contact
with an environment, which in general can be made of
bosons\cite{Leggett}, fermions\cite{Yamada}, or localized
spins\cite{Prokofev}. At the end of the evolution, the properties of
the system are extracted by averaging over the environmental degrees
of freedom.

Significant progress has been made in understanding the effect of
environment on quantum coherence and quantum
computation\cite{Falci,Shnirman,Stamp}. There is evidence that
certain quantum computational models are less sensitive to loss of
phase coherence than others. One of those, is adiabatic quantum
computation (AQC)\cite{Childs,Roland2,Farhi}. In AQC, the system
starts in a known ground state of a Hamiltonian $H_i$. The
Hamiltonian then slowly evolves to a final Hamiltonian $H_f$, whose
ground state encodes the solution to the problem. In the absence of
an environment, if the evolution is slow enough, the system will end
up in the ground state of $H_f$ with probability close to one. The
total time dependent system Hamiltonian is usually written as a
linear interpolation
\be
 H_S(t) = [1-s(t)] H_i + s(t) H_f, \label{HS}
\ee
where $s(t)\in [0,1]$ is a monotonic function of time $t$.

In the absence of an environment, the main source of error for AQC
is non-adiabatic transition out of the ground state. In the small
gap regime, it is closely related to the Landau-Zener (LZ)
transitions\cite{Landau}, which occur when two energy levels pass an
avoided crossing (anticrossing). In the original LZ problem, a
2-level system evolves via a Hamiltonian $H{=}-(\Delta\tau^x {+} \nu
t \tau^z)/2$, where $\tau^{x,z}$ are the Pauli matrices. The
spectrum of this system has an anticrossing at $t{=}0$ with gap
$\Delta$. If at $t{\to}{-}\infty$ the system is in its ground state,
the probability of finding it in its excited state at time
$t{\to}\infty$ is {\em exactly} given by\cite{Landau}
 \be
 P_{\rm LZ}= e^{-\pi \Delta^2/2\nu}. \label{PLZ}
 \ee

In AQC, the system has many energy levels, and the running time is
finite; therefore the above problem does not directly apply.
However, when the gap $g_m$ between the first two levels is much
smaller than other relevant energy scales, the transition
probability is given by \eqref{PLZ} to a very good approximation. In
such a case, $\nu \propto 1/t_f$, where $t_f$ is the running time.
In order to have small excitation probability, one needs a long
$t_f$ ($\propto 1/g_m^2$). Therefore, problems with small gap are
the hardest to solve using AQC.

Recently, there has been compelling evidence that localized two
level systems play an important role in removing phase coherence in
solid state quantum systems\cite{Shnirman,Simmonds,Martinis,Faoro}.
Such 2-level systems can be nuclear spins\cite{Stamp}, magnetic
impurities, or other 2-level fluctuators\cite{Faoro,Koch07}. The
effect of a bosonic environment on LZ transition has been studied
extensively\cite{Ao,Kayanuma,Wubs}, and some studies have been done
for a spin environment\cite{sinitsyn}. In this article, we study a
spin environment focusing only on small gap problems, relevant for
AQC.

\section{Hamiltonian}

As usual we write the total Hamiltonian as $H=H_S+H_B+H_\text{int}$,
which includes system, bath, and interaction Hamiltonians
respectively. If the minimum gap $g_m$ is much smaller than the
separation of the two crossing levels from the other energy levels,
then the slow evolution of the system close to the anticrossing will
be restricted only to those two levels. Using a new coordinate
$\epsilon = 2E(s{-}s^*)$, where $E$ is an energy scale
characterizing the anticrossing and $s^*$ is its position, one can
write a two-state Hamiltonian:
\be H_S =-(\epsilon \tau_z + g_m \tau_x )/2,
 \label{H2L}
\ee
and the gap between the first two states is well approximated by
 \be
 g = \sqrt{\epsilon^2 + g_m^2}. \label{gt}
 \ee
with $\tau_{x,z}$ being the Pauli matrices in the two-state subspace
of the central system. We only focus on linear time sweep for which
$\epsilon = \nu t$ with $\nu=2E/t_f$. In this case, the two-state
problem becomes a LZ problem.

We incorporate the environment via the Hamiltonians
 \be \label{totalH}
 H_B=-\frac{1}{2}\sum_{j=1}^{m}\bm{ B}_j\cdot \bm{\sigma}_j, \qquad
 H_\text{int} = -\frac{1}{2}\sum_{j=1}^{m}
 J_j \tau^z\sigma_j^{z}.
 \ee
Here, $\sigma^{x,z}$ are Pauli matrices for the environmental spins,
$\bm{B}_j{=}[B_j^x,B_j^y,B_j^z]$ is a local field applied to the
$j$-th environmental spin, $J_j$ is the coupling coefficient between
$j$-th spin and the system, and $m$ is the number of environmental
spins. Notice that we only take into account longitudinal coupling
between the effective two-state system ($\tau^z$) and the
environmental spins ($\sigma_j^z$). As is shown in the appendix, in
the case of adiabatic Grover search problem \cite{Roland}, a general
coupling between individual qubits within the system and the
environmental spins leads to only a longitudinal coupling between
the effective two-state system at the anticrossing and the
environmental spins in the limit of large number of qubits. The same
behavior is however expected for more general Hamiltonians because
transverse coupling to the environment will cause large relaxation
to the ground state, enabling us to solve the problem efficiently by
relaying only on classical relaxation (i.e., classical annealing)
which is not expected.

\begin{figure}[pb]
    \centering
        \includegraphics[width=9cm]{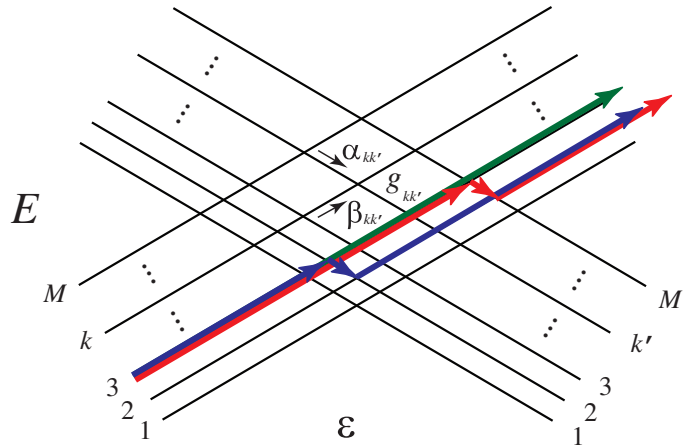}
    \caption{Energy spectrum of the system plus the environment.
    The arrows show paths corresponding to excitation of the system.
    The green line shows a path in the fast passage limit. The red and blue lines show two
    interfering paths. }
    \label{fig1}
\end{figure}

In general, the effect of the environment is to split (broaden) each
energy level into $M{=}2^m$ levels, thereby splitting the original
anticrossing into $M^2$ avoided crossings, as shown in
Fig.~\ref{fig1}. Since we are interested in small $g_m$, we use a
perturbative approach: $H {=} H_0 {+} H'$, with the perturbation
Hamiltonian $H'{=} -g_m \tau^x/2$ and the unperturbed Hamiltonian
$H_0$ which constitutes all the terms in $H$ except $H'$. To
diagonalize $H_0$, we introduce a unitary transformation U by
\ba
\begin{aligned}
 U =& \prod_{j=1}^m \frac{1}{2}\left[ a_j
 + i b_j ( B_j^x\sigma_j^y
 - B_j^y\sigma_j^x)/\sqrt{( B_j^x)^2+( B_j^y)^2}
  \right], \nn
 & a_j = (\cos\theta_j^+ + \cos\theta_j^-)\mathbb{I}
 + (\cos\theta_j^+ - \cos\theta_j^-)\tau^z,  \nn
 & b_j = (\sin\theta_j^+ - \sin\theta_j^-)\mathbb{I}
 + (\sin\theta_j^+ + \sin\theta_j^-)\tau^z, \nn
 & \theta_j^\pm = \arccos \sqrt{(\Omega_j^\pm\pm
 B_j^z+J_j)/2\Omega_j^\pm}, \nn
 & \Omega_j^\pm =
 \sqrt{( B_j^x)^2+( B_j^y)^2+( B_j^z\pm J_j)^2}. \nonumber
\end{aligned}
 \ea
After the transformation, we find
 \ba
 UH_0U^\dagger &=& \sum_{z,s_1,...,s_m}E_{zs_1...s_m}
 |zs_1...s_m\rangle \langle zs_1...s_m|,\\
 UH'U^\dagger &=& -\frac{1}{2}g_m
 \tau^x\prod_{j=1}^m \left[ \cos \theta_j - i\tau^z \sin \theta_j \frac{B_j^x\sigma_j^y - B_j^y\sigma_j^x}
 {\sqrt{( B_j^x)^2 + (B_j^y)^2}} \right], \\
 E_{zs_1...s_m} &=& -\frac{1}{2} \left\{ \epsilon(-1)^z \right.
 + \frac{1}{2}\sum_{j=1}^m(-1)^{s_j} [ (\Omega_j^+-\Omega_j^-)
 \left. +(-1)^z(\Omega_j^++\Omega_j^-)] \right\} \label{energy},
 \ea
where $z$ and $s_j$ are eigenvalues of ($\mathbb{I} - \tau^z$)/2 and
($\mathbb{I} - \sigma_j^z$)/2 respectively, and $\theta_j \equiv
\theta_j^+ +\theta_j^-$.

Let $|z,k\rangle$ denote the state of the system plus the
environment when the environment is in its $k$-th energy level. We
use degenerate perturbation theory to find the gap $g_{kk'}$ at the
point where $|0,k\rangle$ and $|1,k'\rangle$ cross (see
Fig.~\ref{fig1}). Assuming that $H_0$ has no other degeneracies
except at the crossings, we find
 \ba
 g_{kk'} = g_m \prod_{j=1}^m \left| \cos\theta_j \delta_{s_js'_j} 
 + \sin\theta_j (1-\delta_{s_js'_j}) \right|, \label{gkk}
 \ea
where $\delta_{s_js'_j}$ is the Kronecker delta function. Now for
any $k$, one can easily show that
 \ba
\begin{aligned}
 \sum_{k'=1}^{M} g_{kk'}^2 &= g_m^2 \prod_{j=1}^m
 \sum_{s_j'} \left[ \cos^2 \theta_j
 \delta_{s_js'_j} + \sin^2 \theta_j (1-\delta_{s_js'_j}) \right] \\
 &= g_m^2. \label{gapsum}
\end{aligned}
 \ea
This is a very important sum rule which will be used in our analysis
repeatedly.

\section{Environment at zero temperature}

While the intermediate gaps (\ref{gkk}) are proportional to $g_m$,
their separation is independent of $g_m$. Therefore, in the limit
$g_m \to 0$, there is always a regime in which the intermediate gaps
can be considered separated enough to justify using LZ formula,
$p_{kk'}=e^{-\pi g_{kk'}^2/2\nu}$, for the probability of transition
at the $kk'$-th anticrossing. One immediately finds
 \be
 \prod_{k'=1}^{M} p_{kk'} =
 \exp \left\{-\sum_{k'} \pi g_{kk'}^2/2\nu \right\}
 = e^{-\pi g_m^2/2\nu} = P_{\rm LZ},
 \label{Pprod}
 \ee
where $P_{\rm LZ}$ is given by \eqref{PLZ}. Equation \eqref{Pprod}
has an important consequence when the environment is initially in
its ground state ($k{=}1$). The probability of ending up in the
final excited state is the probability that all intermediate LZ
excitations occur (straight path in Fig.~\ref{fig1}): $P_{ef} {=}
\prod_{k=1}^{M} p_{1k} {=} P_{\rm LZ}$. Hence, {\em a zero
temperature spin environment does not affect LZ transition
probability}. Similar results have also been reported for an
environment made of uncoupled harmonic oscillators\cite{Ao,Wubs}.

\section{Environment at finite temperature}

We now discuss the effect of an excited (equilibrium or
nonequilibrium) environment. Let us assume that the system plus the
environment starts from its $i$-th level. A straight path (e.g.,
green path in Fig.~\ref{fig1}) still yields $P_{\rm LZ}$ because of
\eqref{Pprod}. But now there are many other paths to excite the
system, which in general may interfere with each other (e.g., blue
and red paths in Fig.~\ref{fig1}). Those paths, however, can only
add to the probability. Such property also exists if the environment
is initially in a mixed state. Therefore, {\em an initially excited
(incoherent) environment will always increase the LZ transition
probability}. An important question now is to what extent the final
ground state probability $P_{gf}$ will be suppressed as $m$ grows.
If $P_{gf}{\to}0$ as $m{\to}\infty$, then AQC will be impossible in
the presence of such an environment. Fortunately, as we shall show
below, this is not the case.

Phase coherence plays a role when different paths that start form a
common level rejoin in another level and therefore interfere with
each other. The phase that the wave function picks along each path
is the time integral of the eigenenergy along that path. For very
fast evolutions, all the paths come with the same phase (i.e., $\sim
0$). At slower evolutions, different paths will have different
phases. Because of natural randomness in the parameters, the
relative phases of these paths will be randomized and therefore will
eliminate the effect of coherence. Thus, phase coherence is expected
to have negligible effect at long times and large number of paths
(i.e., large $m$). This is justified by our numerical simulations
already at $m=4$ (see Fig.~\ref{fig3}).

In practice, the environmental spins also are subject to thermal
fluctuations, which affect the system via fluctuating local fields
${\bf B}_j$. This effect can be modeled by coupling the spins to an
additional (e.g. bosonic) environment which can cause additional
dephasing between the energy levels making our incoherent approach
even more relevant. Such an environment does not change LZ
probabilities $p_{ij}$ in the fast passage limit ($p_{ij}{\approx}
1$),\cite{Ao} which is always the case when $m{\to}\infty$. Here, we
also neglect any relaxation effect as it can only reduce the
excitation probability, i.e., enhance success probability in AQC.
Hence, our calculation provides a lower bound for the AQC success
probability.

Let $\alpha_{ij}$ and $\beta_{ij}$ be the probabilities for the
upper and lower energy states just before traversing the $ij$-th
anticrossing, respectively (see Fig.~\ref{fig1}). We can write
\be \label{pde} \begin{aligned} &\alpha_{(i-1)j} =
p_{ij}\alpha_{ij}+ (1-p_{ij})\beta_{ij}, \\& \beta_{i(j+1)} =
p_{ij}\beta_{ij} + (1-p_{ij})\alpha_{ij}, \end{aligned}\ee
with boundary conditions: $\alpha_{Mj} {=} P_{ei}^{(j)}$,
$\alpha_{0j} {=} P_{gf}^{(j)}$, $\beta_{i1} {=} P_{gi}^{(i)}$, and
$\beta_{i(M+1)} {=} P_{ef}^{(i)}$. Here $P_{gi}^{(j)}$ and
$P_{ei}^{(j)}$ ($P_{gf}^{(j)}$ and $P_{ef}^{(j)}$) are the initial
(final) probabilities of finding the total system in $|0,j\rangle$
and $|1,j\rangle$, respectively. This has been studied by switching
to continuous variables by Ref.~\refcite{sinitsyn}. Here we instead
solve the discrete relation while keeping the constraint
\eqref{Pprod}. From \eqref{pde} we get
 \ba
 \bm\alpha_{i-1} = P_{gi}^{(i)}\bm\mu^{(i)} + \bm
 L^{(i)}\bm\alpha_i, \nn
 \bm\beta_{j+1} = P_{ei}^{(j)}\bm\eta^{(j)} + \bm
 U^{(j)}\bm\beta_j,
 \ea
with vectors $\bm\alpha_i {=} [\alpha_{i1},...,\alpha_{iM}]^T$,
$\bm\beta_j {=} [\beta_{1j},...,\beta_{Mj}]^T$, $\mu_j^{(i)} {=}
(1-p_{ij})\prod_{l=1}^{j-1}p_{il}$,\ \ $\eta_i^{(j)} {=}
(1-p_{ij})\prod_{l=i+1}^{M}p_{lj}$, and
\ba \label{Li} L_{jk}^{(i)} = \left\{
\begin{array}{ll}
p_{ij} & k = j\\
(1-p_{ik})(1-p_{ij})\prod_{l=k+1}^{j-1}{p_{il}} & j > k\\
0 & k > j
\end{array}\right. , \nn
\label{Uj} U_{ik}^{(j)} = \left\{
\begin{array}{ll}
p_{ij} & i = k\\
(1-p_{kj})(1-p_{ij})\prod_{l=i+1}^{k-1}{p_{lj}} & k > i\\
0 & i > k
\end{array}\right. . \nonumber
\ea

We define the vectors $\bm{P_{gi}}$ and $\bm{P_{ei}}$ ($\bm{P_{gf}}$
and $\bm{P_{ef}}$) to represent the initial (final) ground and
excited state distribution, respectively. Here, we only consider the
case where the central system is initialized in its ground state,
i.e., $\bm{P_{ei}}=\bm{0}$, for which we have
 \be
 \bm{P_{ef}} = \bm{\Lambda} \bm{P_{gi}}, \quad \text{with  }
 \bm\Lambda = \bm U^{(M)}\bm U^{(M-1)} \dots \bm U^{(1)}. \label{incohere}
 \ee
%
The final excited state probability $P_{ef}$ is found by summing
over all elements of $\bm{P_{ef}}$, which is equivalent to tracing
over the environmental degrees of freedom.

Naturally, it is very difficult to find an analytical solution for
the general case. Here, we consider a special case where all the
gaps have the same size, $g_{ij}{=}g_m/\sqrt{M}$, hence
$p_{ij}=p=P_{\rm LZ}^{1/M}$. Such a situation occurs when $|J_j| =
|\bm B_j|$. We find $\bm\Lambda=\bm U^M$, where
\be \label{U} U_{ij} = \left\{
\begin{array}{ll}
p & i = j\\
p^{j-i-1}(1-p)^2 & i < j \\
0 & i > j
\end{array}\right. .
\ee
Although $\bm{U}$ is not diagonalizable, we can still compute the
$M$-th power using its Jordan decomposition. Using induction, it can
be shown that $\bm\Lambda {=} \bm{T}\bm{J^M}\bm{T^{-1}}$, where
\ba J^M_{ij} &=& \left\{
\begin{array}{ll}
p^{M-j+i}{M \choose j-i} & i \leq j \\
0 & i > j
\end{array} \right., \nn
T_{ij} &=& \left\{
\begin{array}{ll}
1 &  j = i = 1 \\
(-p)^{j-i}(1-p)^{-2(j-1)}{j-2 \choose j-i} & j \geq i \geq 2\\
0 & \text{ else }
\end{array}\right. , \nn
T^{-1}_{ij} &=& \left\{
\begin{array}{ll}
1 &  j = i =1 \\
p^{j-i}(1-p)^{2(i-1)}{j-2 \choose j-i} & j \geq i \geq 2\\
0 & \text{ else }
\end{array}\right.. \nonumber
\ea
After some manipulation, the final excitation probability becomes
 \ba
 P_{ef} &=& \sum_{j=1}^{M} v_j P_{gi}^{(j)}, \label{Pef} \\
 v_1 &=& p^M = P_{\rm LZ}, \nn
 v_{j>1} &=& p^{M+j-2}\sum_{t=0}^{j-2}{j{-}2
 \choose t} \left[pq^{2(t+1)}{M \choose t{+}1}
 +\sum_{s=0}^t q^{t+s} {M \choose s}\right], \label{vj}
 \ea
where $q\equiv (1-p)/p$. Indeed, $v_j$ is the excitation probability
when the environment is initially in the $j$-th state
($P_{gi}^{(j)}=1$). Notice that for $P_{\rm LZ} \gtrsim e^{-M}$, $p
\approx 1$ and hence $0<q \ll 1$.

\begin{figure}[bh]
    \centering
        \includegraphics[width=8cm]{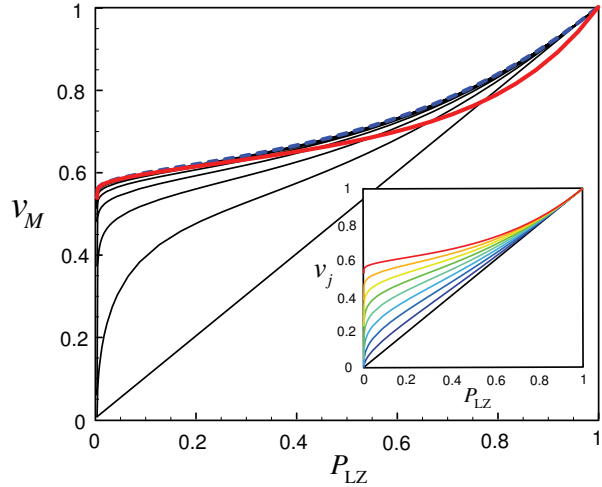}
    \caption{$v_M$ as a function of $P_{\rm LZ}$ ($\equiv e^{-\pi g_m^2/2\nu}$).
    The thin (black) lines are
    obtained using \eqref{vj} with $m=0,1,...,7$, ordered from bottom
    to top. The thick dashed (blue) and solid (red) lines are exact and approximate
    ($M{\to}\infty$) solutions using \eqref{largeM} and \eqref{erf},
    respectively. Inset: $v_j$ vs $P_{\rm LZ}$ for
    $m=7$ and, from bottom to top, $j{=}1$,16,32,48,64,80,96,112, and 128.}
    \label{fig2}
\end{figure}

The inset of Fig.~\ref{fig2} shows $v_j$ vs $P_{\rm LZ}$ for
different values of $j$, when $m{=}7$. The initial rapid rise
happens when $p$ ($=P_{\rm LZ}^{1/M}$) changes from 0 to $\sim$1,
thus becomes extremely sharp at large $m$. Notice that $v_j$
monotonically increases as $j$ grows. This reflects the fact that
higher energy levels have more paths available to them for
excitation. Maximum excitation happens when the system starts in the
highest energy level (i.e., $P_{gi}^{(M)}{=}1$ for which $P_{ef} {=}
v_{M}$), hence $v_M$ gives an upper bound for the excitation
probability ($P_{ef} \leq v_M$). Figure~\ref{fig2} displays $v_M$ as
a function of $P_{\rm LZ}$. As $m$ grows, $P_{ef}$ deviates from
$P_{\rm LZ}$ ($m{=}0$ case). The curves, however, saturate at $m
\gtrsim 6$. This is important because it shows that there is always
a nonzero ground state probability even when $m{\to}\infty$. All
curves asymptotically join the $m{=}0$ curve as $P_{\rm LZ}{\to}1$,
indicating that in the fast passage limit the environment does not
affect the excitation probability.

To find an analytical form for $v_M$ in the large $M$ limit, we use
(\ref{vj}) and write $v_M {=} S_0 {+} S_1$, where
 \ba
 S_0 &=& p^{2M-1}\sum_{t=0}^{M{-}2}q^{2(t+1)}
 {M{-}2 \choose t}{M \choose t{+}1}, \nn
 S_1 &=& p^{2M-2}\sum_{t=0}^{M{-}2}
 q^{t}{M{-}2 \choose t} \sum_{s=0}^tq^{s}{M \choose s}.
 \ea
Using ${M \choose s} = {M{-}2 \choose s} + 2{M{-}2 \choose s{-}1} +
{M{-}2 \choose s{-}2},$ we have
 \ba
 S_1 &=& \frac{1}{2}\left[1 + p^{2M-2}\left(2-p^{-2}\right)W_M \right] - R_M,\nn
 W_M &=& \sum_{t=0}^{M-2}{M{-}2 \choose t}^2 q^{2t}, \nn
 R_M &=& p^{2M-2} \sum_{t=0}^{M-2}{M{-}2 \choose t}{M{-}2 \choose
 t{-}1}q^{2t+1}. \nonumber
 \ea
Now, using
 \ba
 \label{idenBound} \sum_{l=0}^r{m \choose l}{n \choose r{-}l}x^{2l}
 = \frac{1}{2\pi i}\oint_C \frac{(1+z)^n(1+x^2z)^m}{z^{r+1}}
 dz < x^{r-n}(1+x)^{m+n},
 \ea
which holds for $x{>}0$ and any closed contour $C$ around the
origin, one can show that $S_0 {<} pq {\ll} 1$ and $R_M {<} (pq)^2
{\ll} 1$. Choosing an appropriate contour, \eqref{idenBound} yields
 \be \label{wIntegral}
 W_M = \frac{1}{2\pi}\int_{-\pi}^{\pi}\left(1+2q
 \cos\theta+q^2\right)^{M-2} d\theta.
 \ee
As $M \rightarrow \infty$, $\left(1+2q\cos\theta\right)^{M}
\rightarrow P_{\rm LZ}^{-2\cos\theta}$. Hence
 \be
 v_\infty \equiv \lim_{M\to\infty} v_{M} =
 \frac{1}{2}\left(1+\int_{-\pi}^{\pi}\frac{d\theta}{2\pi}P_{\rm
 LZ}^{2(1-\cos\theta)}\right). \label{largeM}
 \ee
For $P_{\rm LZ} \ll 1$, we can further approximate the integrand as
a gaussian to get
 \ba
 v_{\infty} \approx 1/2 + \rm{erf}(\pi \sqrt{-\ln P_{\rm
 LZ}})/4\sqrt{-\pi \ln P_{\rm LZ}}. \label{erf}
 \ea
which gives $v_\infty{\approx}1/2$ in the adiabatic limit ($P_{\rm
LZ}{\to}0$). Equations \eqref{largeM} and \eqref{erf} are plotted in
Fig.~\ref{fig2}. Equation \eqref{erf} provides a good approximation
for $v_\infty$ for a wide range of $P_{\rm LZ}$. This has a very
important implication for AQC. It shows that there is always a
nonzero probability of success bounded from below by $1{-}v_\infty$.
{\em In the adiabatic regime} ($P_{\rm LZ}{\ll}1$), {\em an
environment of the sort modeled here can only suppress the
probability of success by a factor close to 1/2}.

\section{Numerical simulation}

\begin{figure}[bh]
    \centering
        \includegraphics[width=8cm]{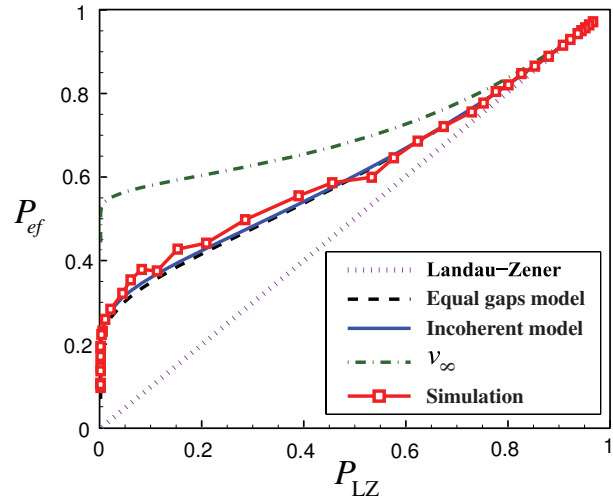}
    \caption{Final excitation probability for completely coherent
    evolution and the approximate solutions. We assumed equally
    populated initial environmental states ($P_{gi}^{(j)}{=}1/M$).
    For the ``incoherent'', ``equal gap'', and ``$v_\infty$'' curves
    we used \eqref{incohere}, \eqref{Pef}, and \eqref{vj} respectively.
    Parameters are: $g_m${=}0.002,
    $[B^x_j,B^z_j]{=}[0.08,-0.04]$, $[-0.07,0.10]$, $[0.15,-0.11]$,
    $[-0.23,0.06]$, $B^y_j{=}0$,
     and $J_j {=} 0.148$ for all $m{=}4$ environmental spins.}
    \label{fig3}
\end{figure}

We have also performed fully coherent simulations, using the
Liouville equation \cite{Blum}
\be
 \dot \rho (t) = -i [H(t), \rho(t)], \label{Liouville}
\ee
for the density matrix $\rho$ of the system plus the environment.
For a case with 4 environmental spins, plotted in Fig.~\ref{fig3},
the coherent simulation shows only small fluctuations around the
incoherent calculation, and also is very close to the equal gap
calculation using \eqref{Pef}. The figure also confirms that
$v_\infty$ provides an upper bound for the excitation probability,
as we expected. We have repeated this simulation for many other
instances and obtained similar behavior.

\section{Conclusions}

We have shown that a $T{=}0$ noninteracting spin environment does
not affect the transition probability in adiabatic quantum
computation in the small gap regime. An excited (equilibrium or
nonequilibrium) environment, on the other hand, will increase the
excitation probability. We found that phase coherence does not play
an important role in the excitation probability. Using an incoherent
model with equal intermediate gaps, we identified a nonzero lower
bound for the ground state probability. In the adiabatic limit, such
a lower bound is ${\sim} 1/2$. AQC therefore is possible in the
presence of such an environment.

\section*{Acknowledgements}

The authors are grateful to  A.J.~Berkley, J.D.~Biamonte, R.~Harris,
G. Martin, G.~Rose, P.C.E.~Stamp, C.J.S.~Truncik, and M.~Wubs for
fruitful discussions.

\appendix

\section{Adiabatic Grover search problem}

In this appendix, we systematically derive the effective two-state
Hamiltonian for the case of adiabatic Grover search (AGS) algorithm.
In AGS, as defined by Roland and Cerf \cite{Roland}, the initial and
final Hamiltonians are
\ba
 \frac{H_i}{E} = \mathbb{I} - |+ \rangle \langle +|
 , \qquad
 \frac{H_f}{E} = \mathbb{I} - |\alpha\rangle \langle \alpha|.
 \label{Htoy}
\ea
where $\mathbb{I}$ is the unity matrix, $|\alpha \rangle$ is the
marked state, and
 \be
 |+\rangle =\frac{1}{\sqrt{N}}\sum_{l}|l\rangle.
 \ee
with $|l\rangle$ representing states in the computation basis. Let
us define a new state
 \be
 |\bar{\alpha}\rangle =\frac{1}{\sqrt{N-1}}\sum_{l\neq \alpha}|l\rangle.
 \ee
which is orthogonal to $|\alpha\rangle$. It is easy to see that
$|+\rangle=(\sqrt{N-1}|\bar{\alpha}\rangle+|\alpha\rangle)/\sqrt{N}$.
The Hamiltonians (\ref{Htoy}) can therefore be written completely in
the subspace made of $|m\rangle$ and $|\bar{m}\rangle$:
\ba
 {H_i / E} = \mathbb{I} -
 \frac{1}{N} \left( \begin{array}{cc}
 1 & \sqrt{N-1} \\
 \sqrt{N-1} & N-1
 \end{array}
 \right), \qquad
 {H_f / E} = \mathbb{I} -
 \left( \begin{array}{cccc}
 1&0\\
 0&0\\
 \end{array}
 \right).
\ea
Using the Pauli matrices $\tau^{x,z}$ in the new 2-state subspace,
\be
 H_S = (1-s)H_i+sH_f = - \frac{1}{2} (\tilde{\epsilon} \tau^z
 + \tilde{g}_m \tau^x), \label{HAGS}
\ee
where
\be
 \tilde{\epsilon} = \frac{E+(N-1)\epsilon}{N}, \qquad
 \tilde{g}_m = \frac{\sqrt{N-1}}{N}(E-\epsilon), \label{gtilde}
\ee
and $\epsilon=E(2s-1)$. We have thrown away trivial terms
proportional to $\mathbb{I}$. In the limit of large $N$, Hamiltonian
(\ref{HAGS}) reduces to (\ref{H2L}) with $g_m=E/\sqrt{N}$.

Let us now introduce interaction between system qubits and
environmental spins via the most general Hamiltonian
 \ba
 H_{\rm int} = - \sum_{\alpha,\beta=x,y,z}\sum_{i=1}^n \sum_{j=1}^m
 J^{\alpha\beta}_{ij}\widetilde{\sigma}_i^\alpha \sigma_j^{\beta} \nonumber
 \ea
Here $\widetilde{\sigma}_i$ and $\sigma_j$ denote Pauli matrices
that act on the $i$-th qubit and $j$-th environmental spin,
respectively, and $m$ is the number of the spins. In the subspace
made of $|\alpha\rangle$ and $|\bar{\alpha}\rangle$, one can write
the reduced interaction Hamiltonian by substituting.
 \ba
 \widetilde{\sigma}_i^x &\to& \frac{1}{N-1}\left( \begin{array}{cc}
 0 & \sqrt{N-1} \\
 \sqrt{N-1} & N-2
 \end{array}
 \right), \nn
 \widetilde{\sigma}_i^y &\to& \frac{1}{N-1}\left( \begin{array}{cc}
 0 & -i\sqrt{N-1} \\
 i\sqrt{N-1} & N-2
 \end{array}
 \right), \nn
 \widetilde{\sigma}_i^z &\to& \left( \begin{array}{cc}
 1 & 0 \\
 0 & -1/(N-1)
 \end{array}
 \right).
 \ea
Or in terms of Pauli matrices $\tau$
 \ba
 \widetilde{\sigma}_i^{x,y} &\to& \frac{N-2}{2(N-1)} (\mathbb{I}-\tau^z)
 + \frac{1}{\sqrt{N-1}} \tau^{x,y} \nn
 \widetilde{\sigma}_i^z &\to& \frac{N-2}{2(N-1)}\mathbb{I} +
 \frac{N}{2(N-1)}\tau^z \label{TLPauli}
 \ea
In the large $N$ limit, one obtains
 \ba
 \widetilde{\sigma}_i^{x,y} \to -\frac{1}{2}\tau^z , \qquad
 \widetilde{\sigma}_i^z \to \frac{1}{2}\tau^z, \label{TLPauliLimit}
 \ea
which yields
 \ba
 H_{\rm int} &=& - \frac{1}{2} \tau^z \sum_{\beta=x,y,z} \sum_{j=1}^m
 J^\beta_j\sigma_j^\beta,
 \ea
where
 \be
 J^\beta_j = \sum_{i=1}^n
 (J^{z\beta}_{ij}-J^{x\beta}_{ij}-J^{y\beta}_{ij}).
 \ee
Since the environmental spins are uncoupled, we can arbitrarily
define the $z$-axis for each spin in such a way to obtain only
longitudinal coupling:
 \be
 H_{\rm int} = - \frac{1}{2}\tau^z \sum_{j=1}^{m} J_j \sigma_j^{z},
 \label{Hint}
 \ee
where
 \be
 J_j = \sqrt{(J^x_j)^2 + (J^y_j)^2 + (J^z_j)^2}.
 \ee

This result is valid more generally than for the spin environment
studied here. In other words, at least for the adiabatic Grover
search problem, any environment that couples to the individual
qubits in a most general way would result in a longitudinal coupling
of the effective two-state system (made of the lowest two energy
states) to the environment. This can be clearly seen from
Eqs.~(\ref{TLPauli}) and (\ref{TLPauliLimit}) that are independent
of the type of environment. Notice that in (\ref{TLPauli}) the terms
involving $\tau^{x,y}$ always appear with a coefficient of
$O(1/\sqrt{N})$ and therefore will vanish in the limit $N\to
\infty$. It physically makes sense because such off-diagonal terms
in the interaction Hamiltonian would cause relaxation between the
two levels with a rate of $O(1/N)$. This means solving the search
problem just based on classical relaxation to the ground state
(i.e., classical annealing) will take a time of $O(N)$ which is the
complexity of the classical search. One therefore would expect that
the same property also persist for problems that are more general
than the adiabatic Grover search; otherwise, e.g., spin glass
problems could be solved efficiently by just letting the system
relax to the ground state.


\begin{thebibliography}{0}
\bibitem{Leggett} A.~J.~Leggett {\it et~al.}, {\it Rev. Mod. Phys.} {\bf 59} (1987) 1.

\bibitem{Yamada} N. Yamada, H.~T.~A.~Sakuma, {\it J. App. Phys.}
{\bf 101} (2007) 09C110.

\bibitem{Prokofev} N.~V. Prokof'ev and P.~C.~E. Stamp, {\it Rep. Prog. Phys.}
{\bf 63} (2000) 669.

\bibitem{Falci} G. Falci {\it et~al.}, {\it Phys. Rev. Lett.}
{\bf 94} (2005) 167002.

\bibitem{Shnirman} A. Shnirman {\it et~al.}, {\it Phys. Rev. Lett}
{\bf 94} (2005) 127002 .

\bibitem{Stamp} A. Morello, P.~C.~E. Stamp, and I.~S. Tupitsyn, {\it Phys. Rev. Lett.}
{\bf 97} (2006) 207206.

\bibitem{Childs} A.~M. Childs, E. Farhi, and J. Preskill, {\it Phys. Rev. A}
{\bf 65} (2001) 012322.

\bibitem{Roland2} J. Roland and N.~J. Cerf {\it Phys. Rev. A}
{\bf 71} (2005) 032330.

\bibitem{Farhi} E. Farhi, J. Goldstone, S. Gutmann, J. Lapan, A. Lundgren,
and D. Preda, {\it Science} {\bf 292} (2001) 472.

\bibitem{Landau} L.~D. Landau and E.~M. Lifshitz,
{\it Quantum mechanics, non-relativistic theory}, (Pergamon Press,
1977).

\bibitem{Simmonds} R.~W. Simmonds {\it et~al.}, {\it Phys. Rev. Lett.}
{\bf 93} (2004) 077003.

\bibitem{Martinis} J.~M. Martinis {\it et~al.}, {\it Phys. Rev. Lett.}
{\it 95} (2005) 210503.

\bibitem{Faoro} L. Faoro {\it et~al.}, {\it Phys. Rev. Lett.}
{\bf 95} (2005) 046805.

\bibitem{Koch07} R.~H. Koch, D.~P. DiVincenzo, and J. Clarke, cond-mat/0702025.

\bibitem{Ao} P. Ao and J. Rammer, {\it Phys. Rev. B} {\bf 43} (1991) 5397.

\bibitem{Kayanuma} Y. Kayanuma and H. Nakayama, {\it Phys. Rev. B}
{\bf 57} (1998) 13099.

\bibitem{Wubs} M. Wubs {\it et~al.}, {\it Phys. Rev. Lett.}
{\bf 97} (2006) 200404.

\bibitem{sinitsyn} N.~A. Sinitsyn and N. Prokof'ev, {\it Phys. Rev. B}
{\bf 67} (2003) 134403.

\bibitem{Blum} K. Blum, {\it Density Matrix Theory and Applications} (Plenum Pub. Corp. 1981).

\bibitem{Roland} J. Roland and N. J. Cerf, {\it Phys. Rev. A}
{\bf 65} (2002) 042308.

\end{thebibliography}
\end{document}